\documentclass[conference]{IEEEtran}
\IEEEoverridecommandlockouts
\usepackage{cite}
\usepackage{amsmath,amssymb,amsfonts}
\usepackage{dsfont}
\usepackage{algorithmic}
\usepackage{graphicx}
\usepackage{textcomp}
\usepackage{xcolor}
\usepackage{pgfplots}
\usepackage{enumitem}
\usetikzlibrary{patterns}
\pgfplotsset{compat=newest}
\pgfplotsset{plot coordinates/math parser=false}
\newlength\figureheight
\newlength\figurewidth

\usepackage{pgfplots}
\pgfplotsset{compat=1.12}
\usepackage{verbatim}
\usetikzlibrary{shapes,arrows}
\usetikzlibrary{positioning}
\usepackage{amsmath,bm,times}
\usepackage{tikz}
\usetikzlibrary{calc}

\usepackage{mathtools}

\DeclarePairedDelimiter{\floor}{\lfloor}{\rfloor}
\usepackage{amsthm}
\usepackage{arydshln}
\usepackage{multirow}
\usepackage{calc}
\usepackage{blkarray}
\usepackage{url}
\theoremstyle{definition}
\newtheorem{theorem}{Theorem}

\newtheorem{lemma}{Lemma}

\newtheorem{example}{Example}
\newtheorem{remark}{Remark}

\usepackage{graphicx,cite}
\usepackage{dblfloatfix}
\usepackage[justification=centering]{caption}
\usepackage{blindtext, graphicx, amsfonts,
	amssymb,multirow,epstopdf}
\usepackage[linesnumbered,ruled]{algorithm2e}
\def\BibTeX{{\rm B\kern-.05em{\sc i\kern-.025em b}\kern-.08em
    T\kern-.1667em\lower.7ex\hbox{E}\kern-.125emX}}

\newcommand{\threevdots}{%
  \vbox{\baselineskip1ex\lineskiplimit0pt%
  \hbox{.}\hbox{.}\hbox{.}}}

\newcommand{\bfZ}{\mathbf{Z}}

\newcommand{\bfH}{\mathbf{H}}

\newcommand{\bfY}{\mathbf{Y}}
\newcommand{\bfW}{\mathbf{W}}
\newcommand{\bfA}{\mathbf{A}}

\newcommand{\bfG}{\mathbf{G}}

\newcommand{\bfx}{\mathbf{x}}

\newcommand{\bfI}{\mathbf{I}}

\begin{document}
\definecolor{mycolor1}{rgb}{0.60000,0.20000,0.00000}%
\definecolor{mycolor2}{rgb}{1.00000,0.00000,1.00000}%
\definecolor{mycolor3}{rgb}{0.00000,0.74902,0.74902}%

\tikzset{%
block1/.style    = {draw, thick, rectangle, minimum height = 1.8em,minimum width = 2.3em},
block2/.style    = {draw, thick, rectangle, minimum height = 2.5em,minimum width = 3.3em},
block3/.style    = {draw, thick, rectangle, minimum height = 2.5em,minimum width = 15em},
block4/.style    = {draw, thick, rectangle, minimum height = 2.5em,minimum width = 9em},
block6/.style    = {draw, thick, rectangle, minimum width = 2em},
block7/.style    = {draw, thick, rectangle, minimum height = 1.8em},
sum/.style      = {draw, circle, node distance = 1cm}, 
  input/.style    = {coordinate}, 
  output/.style   = {coordinate}, 
decision/.style = {diamond, draw, text width=3.5em, text badly centered, node distance=2cm, inner sep=0pt},
block/.style = {rectangle, draw, thick, text centered, minimum height=3em,minimum width = 4 em},
cloud/.style = {draw, ellipse,fill=red!20, node distance=3cm, minimum height=2em},
}
\title{Distributed Matrix-Vector Multiplication: A Convolutional Coding Approach}
\author{\IEEEauthorblockN{Anindya B. Das and Aditya Ramamoorthy}
\IEEEauthorblockA{Department of Electrical and Computer Engineering \\
Iowa State University\\
Ames, IA 50010, U.S.A. \\
\{abd149, adityar\}@iastate.edu}
\thanks{This work was supported in part by the National Science Foundation (NSF) under grant CCF-1718470.}
}

\maketitle
\begin{abstract}

Distributed computing systems are well-known to suffer from the problem of slow or failed nodes; these are referred to as stragglers. Straggler mitigation (for distributed matrix computations) has recently been investigated from the standpoint of erasure coding in several works. In this work we present a strategy for distributed matrix-vector multiplication based on convolutional coding. Our scheme can be decoded using a low-complexity peeling decoder. The recovery process enjoys excellent numerical stability as compared to Reed-Solomon coding based approaches (which exhibit significant problems owing their badly conditioned decoding matrices). Finally, our schemes are better matched to the practically important case of sparse matrix-vector multiplication as compared to many previous schemes. Extensive simulation results corroborate our findings.
\end{abstract}


\section{Introduction}
Distributed computation plays a major role in several problems in optimization and machine learning. For example, large scale gradient descent often requires us to repeatedly calculate matrix-vector products. In high-dimensional problems, time and storage constraints necessitate the splitting of these computations across multiple nodes.



Distributed systems are well known to suffer from the issue of slow or faulty processors, which are referred to as stragglers. Several methods have been developed recently for straggler mitigation by using ideas from erasure coding \cite{dutta2016short, yu2017polynomial, tandon2017gradient}. For example, suppose that we want to compute the product of matrix $\textbf{A} \in \mathbb{R}^{r \times t}$ and $\textbf{x} \in \mathbb{R}^t$ in a distributed fashion. As proposed by \cite{lee2018speeding}, we can first split matrix $\bfA$ into submatrices with equal number of rows as $\bfA^T = \left[\bfA_0^T \;\; \bfA_1^T \right]$, and assign three different worker nodes the jobs of computing $\textbf{A}_0\textbf{x}$, $\textbf{A}_1\textbf{x}$ and $\left(\textbf{A}_0+\textbf{A}_1\right)\textbf{x}$ so that the computational load on each of them is half of the original. It is evident that the master node can recover $\bfA \bfx$, as soon as it receives results from any two workers, i.e., the system is resilient to one straggler. This can be generalized by using Reed-Solomon (RS) code-like ideas. Reference \cite{yu2017polynomial} incorporates similar ideas also for matrix-matrix multiplication. 
In both these cases, the proposed solutions have the property that the master node can recover the final result as soon as it receives the results from any $\tau$ workers; $\tau$ is called the recovery threshold.

While these solutions for distributed matrix-vector multiplication are optimal with respect to the recovery threshold, they neglect certain important issues that exist in practical scenarios. For instance, in many machine learning problems, the matrix $\bfA$ is sparse. RS based aproaches require dense linear combinations of the submatrices of $\bfA$. This may cause the computation time of the submatrix-vector products to go up. 
RS based approaches also suffer from numerical stability issues owing to the high condition number of the corresponding Vandermonde matrices. A high condition number results in the decoded value changing by a large amount even if the computed submatrix-vector products change by a small amount. This is especially relevant in the machine learning context where gradient computations can often be approximate. We note here that this issue with respect to polynomial interpolation is well recognized in the numerical analysis literature \cite{acton1990numerical} and several papers discuss appropriate choices of interpolation points for numerical robustness \cite{berrut2004barycentric}. However, in the straggler mitigation context, we need to be able to perform recovery from any large enough subset of interpolation points. This causes the worst case condition number to be quite bad.


In this work, we present a class of distributed matrix-vector multiplication schemes by leveraging (binary) cross parity check convolutional codes \cite{fuja1989cross}, that were originally proposed for distributed storage systems. In our context, the generator matrices specify the assignment of jobs to the worker nodes.
\subsection{Main Contributions}
\begin{itemize}[wide, labelwidth=!, labelindent=0pt]
\item While the codes in \cite{fuja1989cross} can result in generator matrices that are recursive, we show that our schemes always result in feed-forward encoders. This is important in our setting, because our underlying field of operation is $\mathbb{R}$. Furthermore, we show that in the setting when the number of stragglers is two, then our encoder only has coefficients in $\{-1,0,1\}$. When the number of stragglers is three, we show an upper bound on the absolute value of the coefficients.
\item We demonstrate that our schemes can be decoded at the master node by a low-complexity peeling decoder. This implies that the master node can operate in iterations such that in each iteration, it solves an equation where there is only one unknown.
\item Our experimental results indicate the numerical robustness of our scheme, and also shows its advantage in computation speed when $\bfA$ is sparse.
\end{itemize}

\section{Cross Parity Check Convolutional Codes}

Consider the set of real infinite sequences $\{c_r, c_{r+1}, \dots \}$ for $r \in \mathbb{Z}$ that start at some finite integer index and continue thereafter. These sequences can be treated as elements of the formal Laurent series \cite{fuja1989cross} in indeterminate $D$ with coefficients from $\mathbb{R}$, i.e. $\sum\limits_{i=r}^\infty c_i D^i$. Let us denote the ring of formal Laurent series over $\mathbb{R}$ as $\mathbb{R}((x))$ under the normal addition and multiplication of formal power series. It can be shown that $\mathbb{R}((x))$ forms a field, i.e., each non-zero element of $\mathbb{R}((x))$ has a corresponding inverse.

In this work, we will consider $n$ infinite strips that can be visualized as columns that start at index $r=0$ and continue indefinitely downward. We denote the infinite sequence as $\{c_{0,j}, c_{1,j}, c_{2,j}, \dots \}$ for each column $j$, and so, we can represent the $j$-th strip by the formal series as
\begin{align*}
C_j (D) = \sum\limits_{i=0}^{\infty} c_{i,j} D^i
\end{align*} for $0 \leq j \leq n - 1$. These sequences obey the ``geometric" constraint
\begin{align}
\sum\limits_{j = 0}^{n - 1} c_{i - m j , j} = 0 \text{~~for~} i \geq 0,
\label{eq1}
\end{align} which indicates the lines of slope $m$. For each value of $m = 0, 1, \dots, (n-k-1)$, the sequences sum to zero along the lines of the corresponding slope. The value $c_{i,j} = 0$ if $i < 0$ for any $j$.

Let $CP(n,k)$ denote the set of all sequences\footnote{We will refer to this as the $CP(n,k)$ code} that satisfy the constraints in (\ref{eq1}). This can equivalently be expressed as
%
\begin{align}
\sum\limits_{j = 0}^{n - 1} C_{j}(D) D^{m j} = 0
\label{eq2}
\end{align} for $m = 0, 1, \dots, (n-k-1)$. Let
\begin{align*}
\underline{C}(D) = \left( C_0(D), C_1(D), C_2(D) \dots, C_{n-1}(D) \right) .
\end{align*}
Then, we can express the condition succinctly as
\begin{align*}
 \underline{C}(D) \; \bfH_{n,k}^T(D)= 0 ,
\end{align*} where $\bfH_{n,k}(D)$ is the $ (n-k) \times \, n $  matrix (analogous to a parity-check matrix), which can be obtained from \eqref{eq2} and written as
\begin{align*}
\bfH_{n,k}(D) = \begin{bmatrix}
    1 & 1 & 1 & \dots & 1 \\
	1 & D & D^2 & \dots & D^{(n-1)}\\
	\threevdots & \threevdots & \threevdots &  & \threevdots \\
	1 & D^{(s-1)} & D^{2(s-1)} & \dots & D^{(n-1)(s-1)}
\end{bmatrix}\ .
\end{align*}
Since every $(n-k) \times (n-k)$ submatrix of $H_{n,k}(D)$ is a Vandermonde matrix evaluated at distinct powers of the indeterminate $D$, its determinant will be a non-zero polynomial in $D$ and hence invertible over $\mathbb{R}((x))$. Thus, $\underline{C}(D)$ can be recovered even if any $n-k$ columns are lost.

The key idea underlying our work is that distributed matrix vector multiplication can be embedded into the class of $CP(n,k)$ codes, where a given column $i$ represents (upon appropriate interpretation) the computation assigned to worker node $W_i$, which sequentially processes its assigned jobs from top to bottom. The result $\bfA \bfx$ can be recovered even if any $n-k$ worker nodes fail. Crucially, $\bfA \bfx$ can be decoded using a peeling decoder. This significantly reduces the overall computation load at the master node and provides for a scheme that enjoys excellent numerical stability.

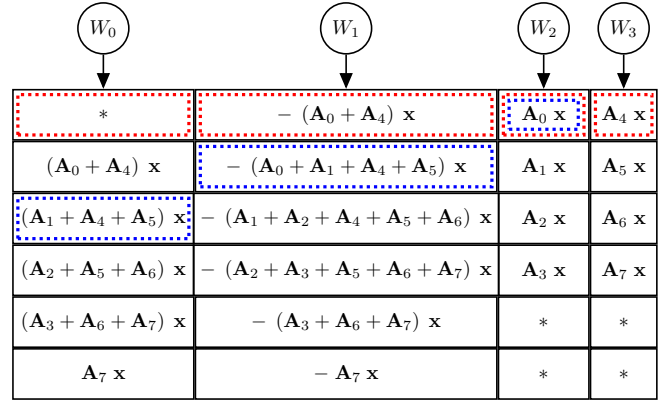
\begin{figure}[t]
\centering

\resizebox{0.99\linewidth}{!}{
\begin{tikzpicture}[auto, thick, node distance=2cm, >=triangle 45]

\draw
	node [block4] (b0) {$*$}
    node [block4, below = 0.0005 cm of b0] (b1) {$\left(\bfA_{0} + \bfA_{4}\right) \; \bfx$}
    node [block4, below = 0.0005 cm of b1] (b2) {$\left(\bfA_{1} + \bfA_{4} + \bfA_{5}\right) \; \bfx$}
    node [block4, below = 0.0005 cm of b2] (b3) {$\left(\bfA_{2} + \bfA_{5} + \bfA_{6}\right) \; \bfx$}
    node [block4, below = 0.0005 cm of b3] (b4) {$\left(\bfA_{3} + \bfA_{6} + \bfA_{7}\right) \; \bfx$}
    node [block4, below = 0.0005 cm of b4] (b41) {$\bfA_{7} \; \bfx$}

    node [block3, right = 0.0005 cm of b0] (b15) {$- \;\left(\bfA_{0} + \bfA_{4}\right) \; \bfx$}
    node [block3, below = 0.0005 cm of b15] (b16) {$- \; \left(\bfA_{0} + \bfA_{1} + \bfA_{4} + \bfA_{5}\right) \; \bfx$}
    node [block3, below = 0.0005 cm of b16] (b17) {$- \; \left(\bfA_{1} + \bfA_{2} + \bfA_{4} + \bfA_{5} + \bfA_{6}\right) \; \bfx$}
    node [block3, below = 0.0005 cm of b17] (b18) {$- \; \left(\bfA_{2} + \bfA_{3} + \bfA_{5} + \bfA_{6} + \bfA_{7}\right) \; \bfx$}
    node [block3, below = 0.0005 cm of b18] (b19) {$- \; \left(\bfA_{3} + \bfA_{6} + \bfA_{7}\right) \; \bfx$}
    node [block3, below = 0.0005 cm of b19] (b191) {$- \; \bfA_{7} \; \bfx$}

    node [block2, minimum width = 4.5 em, right = 0.0005 cm of b15] (b5) {$\bfA_{0} \; \bfx$}
    node [block2, minimum width = 4.5 em, below = 0.0005 cm of b5] (b6) {$\bfA_{1} \; \bfx$}
    node [block2, minimum width = 4.5 em, below = 0.0005 cm of b6] (b7) {$\bfA_{2} \; \bfx$}
    node [block2, minimum width = 4.5 em, below = 0.0005 cm of b7] (b8) {$\bfA_{3} \; \bfx$}
	node [block2, minimum width = 4.5 em, below = 0.0005 cm of b8] (b9) {$*$}
	node [block2, minimum width = 4.5 em, below = 0.0005 cm of b9] (b91) {$*$}

    node [block2, right = 0.0005 cm of b5] (b10) {$\bfA_{4} \; \bfx$}
    node [block2, below = 0.0005 cm of b10] (b11) {$\bfA_{5} \; \bfx$}
    node [block2, below = 0.0005 cm of b11] (b12) {$\bfA_{6} \; \bfx$}
    node [block2, below = 0.0005 cm of b12] (b13) {$\bfA_{7} \; \bfx$}
    node [block2, below = 0.0005 cm of b13] (b14) {$*$}
    node [block2, below = 0.0005 cm of b14] (b141) {$*$}

    node [sum, above = 0.6 cm of b0] (w0){$W_0$}
    node [sum, above = 0.6 cm of b15] (w1){$W_1$}
    node [sum, above = 0.6 cm of b5] (w2){$W_2$}
    node [sum, above = 0.6 cm of b10] (w3){$W_3$}

    ;
    \draw[->](w0) -- node{} (b0);
    \draw[->](w1) -- node{} (b15);
    \draw[->](w2) -- node{} (b5);
    \draw[->](w3) -- node{} (b10);

\draw[ultra thick,red,dotted] ($(b0.north west)-(-0.1,0.1)$)  rectangle ($(b0.south east)-(0.1,-0.1)$);
\draw[ultra thick,red,dotted] ($(b15.north west)-(-0.1,0.1)$)  rectangle ($(b15.south east)-(0.1,-0.1)$);
\draw[ultra thick,red,dotted] ($(b5.north west)-(-0.1,0.1)$)  rectangle ($(b5.south east)-(0.1,-0.1)$);
\draw[ultra thick,red,dotted] ($(b10.north west)-(-0.1,0.1)$)  rectangle ($(b10.south east)-(0.1,-0.1)$);

\draw[ultra thick,blue,dotted] ($(b2.north west)-(-0.1,0.1)$)  rectangle ($(b2.south east)-(0.1,-0.1)$);
\draw[ultra thick,blue,dotted] ($(b16.north west)-(-0.1,0.1)$)  rectangle ($(b16.south east)-(0.1,-0.1)$);
\draw[ultra thick,blue,dotted] ($(b5.north west)-(-0.2,0.2)$)  rectangle ($(b5.south east)-(0.2,-0.2)$);

\end{tikzpicture}
}
\caption{\small Distributed Matrix-vector Multiplication embedded into a $CP(4,2)$ code. The assigned jobs in $W_0$ are downshifted and its first job is denoted by the placeholder *. This is only to make it easy to see that the geometric constraints are satisfied. In reality, $W_0$ will start executing its first job, i.e., $(\bfA_0 + \bfA_4)\bfx$ right away and proceed sequentially downward. Here blue and red dotted blocks indicate examples of two constraint lines with slopes $1$ and $0$, respectively.}
\label{fig2}
\end{figure}

\begin{example}
\label{ex1}
Let $\bfA$ be a large matrix that is split into eight block-rows, $\bfA_0, \bfA_1, \dots, \bfA_{7}$.
Fig. \ref{fig2} shows an example where the distributed computation of $\bfA \bfx$ is embedded into a $CP(4,2)$ code. Each cell in the figure shows the responsibility assigned to the corresponding worker node (from top to bottom). It can be observed that the geometric condition in (\ref{eq1}) is satisfied by the cell contents and that $\bfA \bfx$ can be recovered even if any two of the worker nodes fail. Furthermore, this can be achieved by a peeling decoder with only addition/subtraction operations.
\end{example}


\section{Embedding Matrix-Vector Multiplication into a $CP(n,k)$ code}
\label{sec:embedding}
In this section, we outline the details of our proposed scheme. Towards this end, we first derive the corresponding generator matrix for the $CP(n,k)$ code and show that it can be expressed in feed-forward form. And then we discuss how the distributed computation of $\bfA \bfx$ can be mapped onto a system with $n$ worker nodes using the $CP(n,k)$ code.

Let $s = n-k$ and $\bfY_{a,b}$ be a $a \times b$ matrix such that
\begin{align*}
\bfY_{a,b}(i,j) = D^{i j}; \; \; \; 0\leq i \leq a-1, \;\; 0 \leq j \leq b-1\; ;
\end{align*} and $\Psi_w$ be a $w \times w$ diagonal matrix such that
\begin{align*}
\Psi_{w} = \textrm{diag} \; \begin{bmatrix}
    1 & D & D^2 & \dots & D^{w-1}
\end{bmatrix}\ .
\end{align*}
Thus, we have
\begin{align*}
\bfH_{n,k}^T(D) = \begin{bmatrix}
    \bfY_{s,s} \\
    ---- \\
	\bfY_{k,s} \Psi_s^s
\end{bmatrix}.\
\end{align*} Let the systematic generator matrix be
\begin{align*}
\bfG_{n,k}(D) \; = \; \begin{bmatrix}
    \bfZ \; \; \; \; \bfI_k
\end{bmatrix}\ ,
\end{align*} where $\bfI_k$ is the $k \times k$ identity matrix. Now satisfying $\bfG_{n,k}(D) \; \bfH_{n,k}^T(D) = \textbf{0}$, we obtain

\begin{align}
\label{eq6}
\bfZ \; = \; - \; \bfY_{k,s} \; \Psi_s^s \; \bfY_{s,s}^{-1}.
\end{align} We can express $\bfW_{k,s} = \bfY_{k,s} \; \Psi_s^s$ as
\begin{align*}
\bfW_{k,s}= \begin{bmatrix}
    1 & D^s & D^{2s} & \dots & D^{(s-1)s} \\
	1 & D^{s+1} & D^{2(s+1)} & \dots & D^{(s-1)(s+1)}\\
	\threevdots & \threevdots & \threevdots &  & \threevdots \\
	1 & D^{(s+k-1)} & D^{2(s+k-1)} & \dots & D^{(s-1)(s+k-1)}
\end{bmatrix}.\
\end{align*} Suppose that $\{y_{\ell j}\}$ are the elements of $\bfY_{s,s}^{-1}$ and we define $s$ different polynomials in the indeterminate $\sigma$ as
\begin{align*}
y_j(\sigma) = \sum\limits_{\ell = 0}^{s - 1} \;  y_{\ell j} \; \sigma^\ell
\end{align*} for $0 \leq j \leq s$. Since $\bfY_{s,s} \bfY_{s,s}^{-1} = \textbf{I}$, we can write for any $j$,
\begin{align*}
y_j (D^\ell) = 0 , \; \; \; \; \; \textrm{for} \; \; \ell \neq j ,
\end{align*} which indicates that any $D^\ell$ will be a root of the polynomial $y_j(\sigma)$ if $\ell \neq j$. On the other hand, $y_j (D^j) = 1$ so that
\begin{align*}
y_j(\sigma) = \sum\limits_{\ell = 0}^{s - 1} \;  y_{\ell j} \; \sigma^\ell \; = \; \prod\limits_{\ell = 0, \ell \neq j}^{s-1} \frac{\sigma - D^\ell}{D^j - D^\ell} .
\end{align*} Now taking the product $\bfW_{k,s} \; \bfY_{s,s}^{-1}$ involves evaluating these polynomials at $D^{s+i}$ where $i = 0, \dots, k-1$, and thus we get
\begin{align}
\label{eqZ}
Z_{ij} = \; - \; \prod_{\ell = 0, \ell \neq j}^{s - 1} \; \frac{D^{s+i} - D^{\ell}}{D^j - D^{\ell}}.
\end{align}
It is unclear whether the above expression leads to a recursive or a non-recursive $\bfG_{n,k}$. The following theorem shows that $Z_{ij}$ can be simplified to express it as  a polynomial, i.e., $\bfG_{n,k}$ can be put in feed-forward form.

\begin{theorem}
\label{thm:generator}
Any term $Z_{ij}$, with $0 \leq i < k$ and $0 \leq j < s$, can be written as a finite polynomial in $D$ with integer coefficients.
When $s = 2$, the coefficients of $Z_{ij} \in \{-1,0,1\}$ and when $s=3$ the coefficients of $Z_{ij}$ have absolute value at most $k$.

\end{theorem}

\begin{proof}
From \eqref{eqZ}, we can write
\begin{align*}
& Z_{ij}  = -  \left[ \frac{\left( D^{s+i} - D^{0}\right)\left( D^{s+i} - D^{1}\right) \dots \left( D^{s+i} - D^{j-1}\right)}{\left( D^{j} - D^{0}\right)\left( D^{j} - D^{1}\right) \dots \left( D^{j} - D^{j-1}\right)} \right] \\  \times & \left[ \frac{\left( D^{s+i} - D^{j+1}\right)\left( D^{s+i} - D^{j+2}\right) \dots \left( D^{s+i} - D^{s-1}\right)}{\left( D^{j} - D^{j+1}\right)\left( D^{j} - D^{j+2}\right) \dots \left( D^{j} - D^{s-1}\right)} \right]
\end{align*} which can be written as $Z_{ij} = - \; A_{ij} B_{ij}$. Here
\begin{align*}
A_{ij} &= \frac{\left( D^{x_{ij}+j} - 1\right)\left( D^{x_{ij}+j-1} - 1\right) \dots \left( D^{x_{ij}+1} - 1\right)}{\left( D^{j} - 1\right)\left( D^{j-1} - 1\right) \dots \left( D^{1} - 1\right)} \; ;\\
B_{ij} &= (-1)^{e_{ij}} D^{f_{ij}} \frac{\left( D^{i+y_{ij}} - 1\right)\left( D^{i+y_{ij}-1} - 1\right) \dots \left( D^{i+1} - 1\right)}{\left( D^{1} - 1\right)\left( D^{2} - 1\right) \dots \left( D^{y_{ij}} - 1\right)} \; ;
\end{align*} where $x_{ij} = s + i - j$, $y_{ij} = s - j - 1$, $e_{ij} = s - j - 1$ and $f_{ij} \geq 0 $ as $j < s$. Now consider a term $W$ such that
\begin{align*}
W  = \frac{\left( D^{x+1} - 1\right)\left( D^{x+2} - 1\right) \dots \left( D^{x+y} - 1\right)}{\left( D^{1} - 1\right)\left( D^{2} - 1\right) \dots \left( D^{y} - 1\right)} .
\end{align*} A polynomial of the form $D^q - 1$ can be written as a product of some cyclotomic polynomials \cite{nagell1964introduction}, as
\begin{align*}
D^q - 1 = \prod_{d|q} \Phi_d (D)
\end{align*} where $\Phi_d (D)$ is the $d^{th}$ cyclotomic polynomial. Thus, we can expand the denominator of $W$ as
\begin{align*}
\textrm{den} (W) = \left[ \Phi_1 (D) \right]^{i_1} \times \left[ \Phi_2 (D) \right]^{i_2} \times \dots \times \left[ \Phi_y (D) \right]^{i_y}
\end{align*} where $i_j = \floor{y / j}$ for $j = 1, 2, \dots, y$. The numerator of $W$ is a product of $y$ different polynomials, and the maximum exponent of $D$ in those $y$ polynomials are consecutive $y$ numbers. This indicates that we must have  $\prod\limits_{v=1}^y \left[ \Phi_v (D) \right]^{i_v}$ as a factor of the numerator. But the numerator is, in fact, a product of different cyclotomic polynomials, according to its definition.  Thus, after the cancellation by the denominator and expansion, all the exponents and corresponding coefficients of $D$ in $W$ would be integers. It indicates that the terms $A_{ij}$ and $B_{ij}$ are also finite polynomials of $D$ with integer coefficients, and thus the first part of the proof is complete. The proof regarding the corresponding coefficients of the assigned submatrix-vector block products is given in Appendix \ref{A}.
\end{proof}

\begin{example}
\label{ex2}

When $n=4, k=2$, we can obtain $\bfG_{4,2}(D)$ as
\begin{align}
\label{eq7}
\bfG_{4,2}(D) \; = \; \begin{bmatrix}
 D & - D - 1  & 1 & 0\\
D^2 + D & - D^2 - D - 1 & 0 & 1
\end{bmatrix}\
\end{align}
where we note that all coefficients of the polynomials in $\bfG_{4,2}(D)$ belong to the set $\{-1,0,1\}$.
\end{example}

Next, we discuss the usage of $\bfG_{n,k}$ in a distributed matrix-vector multiplication context. First we partition $\bfA$ row-wise into $\Delta$ (we assume that $k$ divides $\Delta$) equal sized block-rows. These are denoted $\bfA_0, \bfA_1, \bfA_2, \dots, \bfA_{\Delta - 1}$. Let $u_j = \bfA_j \bfx$ where $(0 \leq j \leq \Delta - 1)$. The next step is to form $k$ different polynomials with coefficients from the $u_j$'s. These are given by
\begin{align*}
\tilde{u}_i(D) = u_{iq} + u_{iq+1} D + \dots + u_{(i+1)q \, - \, 1} D^{q - 1}
\end{align*} for $0 \leq i \leq k-1$ where $q = \frac{\Delta}{k}$. Now the submatrices assigned to all the workers are determined by
\begin{align*}
U(D) \; = \; \begin{bmatrix}
 \tilde{u}_0(D) & \tilde{u}_1(D) & \dots & \tilde{u}_{k-1}(D)
\end{bmatrix} \; \; \bfG_{n,k}(D) .
\end{align*} Suppose that each worker node can store at most the equivalent of $\gamma$ fraction of the rows of $\bfA$. Thus, if we assign $\ell_j$ jobs to worker $W_j$, then it needs to satisfy $\frac{\ell_i}{\Delta} \leq \gamma$. The number of jobs assigned to worker node $W_i$, $(0 \leq i \leq n-1)$, depends on the entries of the corresponding column of $\bfG_{n,k}(D)$. 
Let $d_i$ denote the difference between the maximum and the minimum exponent of $D$ in the $i$-th column of $\bfG_{n,k}(D)$, and let $\lambda = \underset{i}{\text{max}} \; d_i$. Then, for satisfying the storage constraint we require

\begin{align}
\label{eqDel}
\frac{\Delta}{k} \; + \; \lambda \; \leq \; \gamma \; \Delta \; \; \; \; \; \; \; \textrm{which leads to} \; \;  \; \; \; \; \Delta \geq \frac{\lambda}{\gamma - \frac{1}{k}}.
\end{align} 

\begin{example}
\label{ex3}
We consider the same scenario as mentioned in Example \ref{ex2} with $n= 4$ workers, and we need to develop a scheme that is resilient to $s = 2$ stragglers, so $k = n - s = 2$. If $\gamma = \frac{3}{4}$, then we can set $\Delta = \frac{2}{\frac{3}{4} - \frac{1}{2}} = 8$, as it is divisible $k$. The relevant polynomials are 
\begin{align*}
&\tilde{u}_0(D) = u_0 + u_1 D + u_2 D^2 + u_3 D^3, \text{~and} \\
&\tilde{u}_1(D) = u_4 + u_5 D + u_6 D^2 + u_7 D^3.
\end{align*}
Forming $U(D) = \begin{bmatrix}
 \overset{\sim}{u_0}(D) & \overset{\sim}{u_1}(D)
\end{bmatrix} \bfG_{4,2}(D)$, we obtain the scheme shown in Fig. \ref{fig2}.
\end{example}

\begin{remark}
The above procedure demonstrates the importance of a feed-forward $\bfG_{n,k}$. Indeed, if $\bfG_{n,k}$ had been recursive, the number of terms in the output would have been infinite.
\end{remark}

\section{Decoding of a $CP(n,k)$-scheme}
A major advantage of our proposed scheme is a low-complexity decoding procedure. Note that our scheme is in one-to-one correspondence with the $CP(n,k)$ code. Hence, we describe the decoding procedure for the $CP(n,k)$ code; the adaptation to recovering $\bfA \bfx$ follows naturally.

Recall that the symbols are denoted by $c_{i,j}$ where $0 \leq i \leq \ell_j - 1$ and $0 \leq j \leq n - 1 $. Let the indices of the straggler nodes be $0 \leq t_0 < t_1 < t_2 < \dots < t_{s-1} \leq n - 1$.

At each step, our decoding process in Algorithm \ref{Alg:dec} exploits the geometric constraints in \eqref{eq1} to identify an equation where there is one unknown; it continues in a systematic fashion until all the unknowns are decoded. In the sequel, we refer to this as decoding in a peeling decoder fashion. For instance, to start decoding symbols from straggler $t_0$, we can use \eqref{eq1} and obtain the constraint for the line with slope $s-1$ as
\begin{align*}
\sum\limits_{j = 0}^{n - 1} c_{i - (s-1) j , j} = 0
\end{align*} which can pass through a symbol $c_{\alpha,t_0}$ in straggler $t_0$. So we have the constraint as
\begin{align}
\label{eq9}
\sum\limits_{j = 0}^{n - 1} c_{(s-1)(t_0 - j) + \alpha, j} = 0
\end{align} In \eqref{eq9}, if $j < t_0$, we assumed that these symbols are known. Now if $j > t_0$, the elements $c_{\alpha,j}$'s are also known until a constraint line passes through $c_{0,t_1}$  with the increase of $j$. Thus, in the extreme case, the line can pass through $c_{-1,t_1}$, so we can set $j = t_1$ and
\begin{align*}
\alpha + (s - 1)(t_0 - j) = -1 \, .
\end{align*} Thus if $0 \leq \alpha \leq (s - 1)(t_1 - t_0) - 1$, we can say that the only unknown in \eqref{eq9} is $c_{\alpha, t_0}$. In this manner, we can obtain the first $(s-1)(t_1 - t_0)$ symbols of straggler $t_0$ in a peeling decoding fashion.

To better understand the behavior of the algorithm, we divide it into phases. We say that the algorithm has finished phase $p$, where $0 \leq p \leq s-2$, if it has only recovered as many symbols as possible in a peeling decoding fashion from the stragglers $t_i$, $i \leq p$, without recovering any symbol from stragglers $t_j$, $j > p$. Let us denote $\eta_{p,y}$ as the number of recovered symbols in a peeling decoding fashion from straggler $t_y$ after finishing phase $p$.

\begin{lemma}
\label{lem3}
\begin{align}
\label{ind}
\eta_{p,y} \; = \begin{cases}
      \; \sum\limits_{i = y}^{p} \; \left( s - 1 - i \right) \left( t_{i+1} - t_i \right) \; , & \textrm{if}  \; \; y \leq p \; ;\\
      \; \; \; \; \; \; \; \; \; \; \; \; \; \; 0 \; , & \textrm{otherwise} \; ;
    \end{cases}
\end{align} for $p < s - 1$.
\end{lemma}
\begin{proof} We are going to prove this by induction.\\
\noindent \textbf{Base case:} After finishing the phase $p = 0$, we recover as many symbols as possible from straggler $t_0$. Thus we have proved earlier in this Section that
\begin{align*}
& \eta_{0,0} \; = \; (s-1)(t_1 - t_0) \; , \; \textrm{and} \\ & \eta_{0,y} \; = \; \; \;  0 \; ,\; \; \; \; \textrm{for} \; \; \; \; y > 0 .
\end{align*}

\noindent \textbf{Inductive step:} Now we assume that after finishing phase $p$, the number of recovered symbols by the stragglers can be represented by
\begin{align*}
\eta_{p,y} \; = \begin{cases}
      \; \sum\limits_{i = y}^{p} \; \left( s - 1 - i \right) \left( t_{i+1} - t_i \right) , & \textrm{if}  \; \; y \leq p \; ;\\
      \; \; \; \; \; \; \; \; \; \; \; \; \; \; 0\; ,  & \textrm{otherwise} \, ;
    \end{cases}
\end{align*} where $ 0 \leq p < s - 2$. Then after finishing phase $p + 1$, to recover symbols from the straggler $t_{p+1}$, we obtain the constraint for the line with slope $(s-p-2)$ as
\begin{align}
\label{eq10}
\sum\limits_{j = 0}^{n - 1} c_{i - (s-p-2) j , j} = 0 ,
\end{align} which can pass through $c_{0,t_{p+1}}$. So we obtain
\begin{align*}
\sum\limits_{j = 0}^{n - 1} c_{(s-p-2)(t_{p+1}-j) , j} = 0 .
\end{align*} For $j < t_{p+1}$, if the line passes through the straggler $t_q$ where $0 \leq q \leq p$, it will pass through the points $c_{(s-p-2)(t_{p+1}-t_q),t_q}$, but from the hypothesis we assumed $\eta_{p,q}$ symbols are known from worker $t_q$ after phase $p$, and
\begin{align*}
& \eta_{p,q} \;  = \;\sum\limits_{i=q}^{p} (s-1-i)(t_{i+1} - t_i) \\
& > \;\sum\limits_{i=q}^{p} (s-p-2)(t_{i+1} - t_i) =  (s-p-2)(t_{p+1}-t_q) \; ,
\end{align*} which indicates that all necessary symbols of the stragglers are already known. And, if $j > t_{p+1}$, then the row index is negative which means that those corresponding elements are zero. So, we can recover $c_{0,t_{p+1}}$ in a peeling decoding fashion.

Now, we move to straggler $t_p$ and want to recover one more symbol, where already we recovered $\eta_{p,p}$ symbols. So, we can find the constraint line with slope $(s-p-1)$,
\begin{align*}
\sum\limits_{j = 0}^{n - 1} c_{i - (s-p-1) j , j} = 0
\end{align*} and to recover one additional symbol from straggler $t_p$, this line needs to pass through $c_{(s-p-1)(t_{p+1} - t_p),t_p}$, so we obtain
\begin{align*}
\sum\limits_{j = 0}^{n - 1} c_{(s-p-1)\left(t_{p+1} - j\right) , j} = 0 .
\end{align*} In this equation, the terms with $j > t_{p+1}$ are actually zero, and the term with $j = t_{p+1}$ is $c_{0,t_{p+1}}$ which is decoded before. Now if $j = t_q < t_{p}$, the values are already known after finishing phase $p$ because
\begin{align*}
& \eta_{p,q} \;  = \;\sum\limits_{i=q}^{p} (s-1-i)(t_{i+1} - t_i) \\
& > \;\sum\limits_{i=q}^{p} (s-p-1)(t_{i+1} - t_i) =  (s-p-1)(t_{p+1}-t_q).
\end{align*} So we can obtain $c_{\eta_{p,p},t_p}$ in a peeling decoding fashion.

Thus we need previously decoded $c_{0,t_{p+1}}$ to decode $c_{\eta_{p,p},t_p}$. Now in a similar way, we can decode all $c_{\eta_{p,q},t_q}$, where $q < p$, each of which would need the previously decoded $c_{\eta_{p,q+1},t_{q+1}}$. It should be noted that we do not need any  symbols from straggler $t_{i}$ to decode $c_{0,t_{p+1}}$, where $i > p+1$. Thus we do not need them to decode $c_{\eta_{p,q},t_q}$ too, $(q \leq p+1)$, because of the constraint lines with increasing slopes. This argument is sketched in Appendix \ref{B}.

Now we repeat the whole process until we require any symbol from straggler $t_{p+2}$ to decode a symbol from the straggler $t_{p+1}$. We assume that we can recover maximum $\nu$ symbols from $t_{p+1}$ without decoding any symbol from the straggler $t_q$, $q > p+1$. So, the constraint line in \eqref{eq10} passes through $c_{\nu, t_{p+1}}$ and $c_{-1, t_{p+2}}$, and then, we can write
\begin{align*}
\nu \; + \; (s - p - 2) (t_{p+1} - t_{p+2}) \; = \; - 1
\end{align*} which leads to $\nu= (s - p - 2) (t_{p+2} - t_{p+1}) - 1$. So we can recover $(s-p-2)(t_{p+2} - t_{p+1})$ symbols from each of the stragglers, $t_0, t_1, \dots, t_{p+1}$. So, we can say
\begin{align*}
\eta_{p+1,y} \; = \begin{cases}
      \; \;\sum\limits_{i=y}^{p+1} (s-1-i)(t_{i+1} - t_i)  , & \textrm{if}  \; \; y \leq p+1 ;\\
      \; \; \; \; \; \; \; \; \; \; \; \; \; \; 0 , & \textrm{otherwise};
    \end{cases}
\end{align*} which proves the inductive step.
\end{proof}
\begin{theorem}
\label{thm:peeling}
The decoding procedure in Algorithm \ref{Alg:dec} allows the recovery of all $c_{i,j}$ where $0 \leq i \leq \ell_j - 1$ and $0 \leq j \leq n - 1$ in a peeling decoding fashion, as long as there are at most $s$ node failures.
\end{theorem}
\begin{proof}
From Lemma \ref{lem3}, we have proved that we can decode $\eta_{s-2,q}$ elements from a straggler $t_q$ after finishing phase $s-2$, for $q < s-1$. Now if we apply the constraint line with slope zero, then we can recover $\eta_{s-2,s-2} = (t_{s-1} - t_{s-2})$ elements from straggler $t_{s-1}$, and similar to the proof of inductive step, applying the other slopes we can recover additional $(t_{s-1} - t_{s-2})$ elements from other stragglers, until all the symbols from a straggler are recovered. If we just continue this process, we can recover all symbols from all the workers in a peeling decoding fashion.
\end{proof}

\begin{example}
\label{DecExp}
In this example, we demonstrate how we can recover $\bfA \bfx$ in a peeling decoding fashion for the same scenario in Fig. \ref{fig2}. In this example, we have four workers, $W_0, W_1, W_2$ and $W_3$, among which we assume that $W_2$ and $W_3$ are stragglers. According to Algorithm \ref{Alg:dec}, we utilize the slope $1$ constraint lines to recover the block products of $W_2$ and slope $0$ constraint lines to recover the block products of $W_3$.
For example, if we utilize the slope $1$ constraint line through the blue dotted blocks in Fig. \ref{fig2}, then we can recover the first block of of $W_2$, which is $\bfA_0 \bfx$. Similarly, if we utilize the slope $0$ constraint line through the red dotted blocks, then we can also recover the first block of $W_3$, which is $\bfA_4 \bfx$. Using this same fashion, we can decode all the submatrix-vector block products from workers $W_2$ and $W_3$, which completes the job.
\end{example}

\begin{algorithm}[t]
	\caption{Decoding of $CP(n,k)$ Scheme}
	\label{Alg:dec}
   \SetKwInOut{Input}{Input}
   \SetKwInOut{Output}{Output}
   \Input{A $CP(n,k)$ scheme to obtain $\bfA \bfx$ and the index of the stragglers}
   Sort the stragglers according to their index as $0 \leq t_0 < t_1 < t_2 < \dots < t_{s-1} \leq n-1 $\\
\For{$j\gets 0$ \KwTo $s - 2$}{
\For{$i\gets 0$ \KwTo $(s-1-j)(t_{j+1} - t_j) - 1$}{
Set $q = j$ ;\\
\While{$q \geq 0$}{
Obtain an additional submatrix-vector block product (until $\ell_{t_q}$) in straggler $t_{q}$ using constraint line with slope $s-1-q$ ; \\
$q \gets q - 1$ ;
   }
   }
   }
\For{$j\gets 0$ \KwTo $\ell_{s-1}-1$}{
Set $q = s-1$ ;\\
\While{$q \geq 0$}{
Obtain an additional submatrix-vector block product (until $\ell_{t_q}$) in straggler $t_{q}$ using constraint line with slope $s-1-q$ ;\\
$q \gets q - 1$ ;
   }
   }
\Output{All block products to obtain $\bfA \bfx$}
\end{algorithm}

\section{Numerical Results}

We consider a scenario where $\bfA$ has dimension $8,000 \times 10,000$ and $\bfx$ is of length $10,000$. Suppose that we have $n = 7$ workers, $\gamma = \frac{3}{10}$ (storage fraction). We require the system to be resilient to $s = 3$ stragglers.

For the RS based approach we can partition $\bfA$ in $\Delta = 10$ parts, and assign three submatrix-vector products to each worker. The evaluation points for the RS code are chosen as real numbers equally spaced in $[-1,1]$; this is a better choice than integers \cite{8528366}. On the other hand, we can embed this problem into a $CP(7,4)$ using our proposed approach. The value of $\lambda$ ({\it cf.} Section \ref{sec:embedding}) in this case is $8$ and $\Delta = \frac{\lambda}{\gamma - \frac{1}{k}} = \frac{8}{\frac{3}{10} - \frac{1}{4}} = 160$. The assignment of jobs to each worker node can be determined by the procedure in Section \ref{sec:embedding}

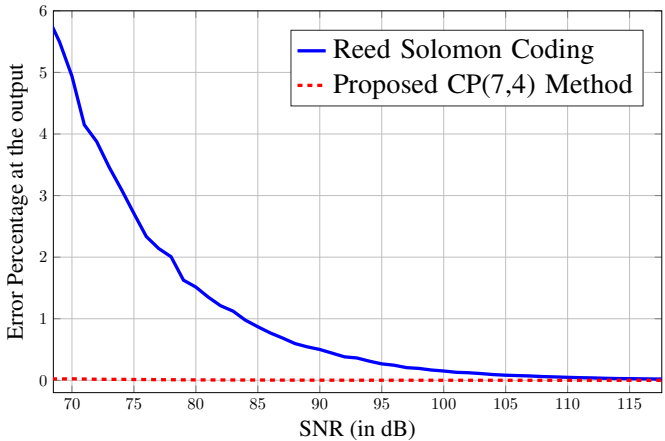
\begin{figure}[t]
\centering
\resizebox{1\linewidth}{!}{
\begin{tikzpicture}

\begin{axis}[%
width=5.1in,
height=3.203in,
at={(2.6in,0.756in)},
scale only axis,
xmin=68.5,
xmax=117.6,
xlabel style={font=\color{white!15!black}, font=\Large},
xlabel={SNR (in dB)},
ymin=-0.2,
ymax=6,
ylabel style={font=\color{white!15!black}, font= \Large},
ylabel={Error Percentage at the output},
axis background/.style={fill=white},
xmajorgrids,
ymajorgrids,
legend style={at={(0.39,0.75)}, nodes={scale=1.7}, anchor=south west, legend cell align=left, align=left, draw=white!15!black}
]
\addplot [color=blue, line width=2.0pt]
  table[row sep=crcr]{%
68	5.92390962802465\\
69	5.49524615333694\\
70	4.93925356522449\\
71	4.14861484345531\\
72	3.87561838241775\\
73	3.46265547424299\\
74	3.09672378656027\\
75	2.71024119387205\\
76	2.3350506721627\\
77	2.14029774686099\\
78	2.00650616136716\\
79	1.62632802259374\\
80	1.51433501034654\\
81	1.35086946569988\\
82	1.21074017807479\\
83	1.12381542062707\\
84	0.976376647327221\\
85	0.869722808968313\\
86	0.769596788248796\\
87	0.686744549792312\\
88	0.596870965778296\\
89	0.54338075624165\\
90	0.500936938316075\\
91	0.440843122439805\\
92	0.381196352321356\\
93	0.362791895958372\\
94	0.311114173103023\\
95	0.267100685638432\\
96	0.245281870481972\\
97	0.206100413759563\\
98	0.193326024079774\\
99	0.167269735006305\\
100	0.151338920973577\\
101	0.130312163214881\\
102	0.123148449746791\\
103	0.109625258884205\\
104	0.093183898917619\\
105	0.0828222457531148\\
106	0.0765853043557554\\
107	0.0704372576243645\\
108	0.0608621760494631\\
109	0.0552151102783731\\
110	0.0483665258119865\\
111	0.0429408148943326\\
112	0.0391760602681063\\
113	0.033358250128903\\
114	0.0285526638215253\\
115	0.0271993878215265\\
116	0.0245035912361556\\
117	0.0217635080715183\\
118	0.0199745237571612\\
119	0.0168451559625786\\
120	0.0149522359429769\\
};
\addlegendentry{Reed Solomon Coding}

\addplot [color=red, dashed, line width=2.0pt]
  table[row sep=crcr]{%
66	0.0351366081697394\\
67	0.0296824895381235\\
68	0.0270306254036843\\
69	0.0216995969785643\\
70	0.0246808122356224\\
71	0.0211112799819795\\
72	0.0174245900144727\\
73	0.0157946938103516\\
74	0.0158618360831681\\
75	0.0130691333293402\\
76	0.0126625937555496\\
77	0.010229452728978\\
78	0.00963151392820885\\
79	0.00761422029174592\\
80	0.00739168489415744\\
81	0.00614774776843978\\
82	0.00580719651552792\\
83	0.00552609070241474\\
84	0.00444797274907162\\
85	0.00520757193926452\\
86	0.00349617369460333\\
87	0.00325154113227767\\
88	0.0029618256155178\\
89	0.00272935784721189\\
90	0.00226301645600045\\
91	0.00218624395866514\\
92	0.00153020519641167\\
93	0.00141322067637255\\
94	0.00171748901488297\\
95	0.00144793703349291\\
96	0.00110421949003718\\
97	0.00117193935049999\\
98	0.000953454691779692\\
99	0.000946496649881478\\
100	0.000748602415528682\\
101	0.000615871911176441\\
102	0.000589075783118173\\
103	0.00052723239744192\\
104	0.000516055113146194\\
105	0.000442487197925173\\
106	0.000333971597774091\\
107	0.000321931345267481\\
108	0.000265015755516296\\
109	0.00021497422628491\\
110	0.000256782078782607\\
111	0.000169244471516602\\
112	0.000203884106721274\\
113	0.00019676533856435\\
114	0.000129881701676641\\
115	0.000183674292273144\\
116	0.000122096317528375\\
117	0.000107005719263706\\
118	8.75584111627382e-05\\
119	7.67586519169811e-05\\
120	7.57439066191114e-05\\
};
\addlegendentry{Proposed CP(7,4) Method}
\end{axis}
\end{tikzpicture}%
}
\caption{\small Comparison between our proposed method and RS coding based method at different noise levels}
\label{fig4}
\end{figure}

In order to test the numerical stability of recovery at the master node, we add white Gaussian noise to each submatrix-vector product computed by the workers. 
Fig. \ref{fig4} presents a comparison of the RS based approach and our proposed approach. The error percentage values at the output are shown for different noise levels. If the correct output is $y$ and the recovered output by the master is $\hat{y}$, then the error percentage is measured as $\frac{||y-\hat{y}||}{||y||} \times 100\%$. We can see that the error percentage for our proposed method is nearly zero, whereas the RS based method has around $5\%$ error even at an SNR = $70$ dB. This is due to the high condition number of the associated real Vandermonde matrix of the RS based method.

Next, we compare the computation times of the two schemes when the matrix $\bfA$ is sparse. 
Consider a system with $n = 5$ workers with $\gamma = \frac{3}{5}$. We choose $\bfA$ (of dimension $12,000 \times 12,000$) to be of limited bandwidth \cite{golub2012matrix}, i.e., it has non-zero values only in the $\beta$- diagonals (diagonals from top-left to bottom-right) and $\beta = -b, -b+1, \dots, -1, 0, 1, \dots, b-1, b$, where $b < 12,000$. Thus the sparsity of the matrix decreases if $b$ increases.

In the RS based approach, we choose $\Delta = 5$ and assign $3$ jobs to each of the workers. For our scheme we embed the computation into a $CP(5,2)$ code. The parameter $\lambda = 4$, and $\Delta = \frac{\lambda}{\gamma - \frac{1}{k}} = 40$. 
Note that both of the schemes are resilient to three stragglers.

\begin{figure}[t]
\centering
\resizebox{1\linewidth}{!}{
\begin{tikzpicture}
\begin{axis}[
width=5in,
height=3.203in,
at={(2.6in,0.852in)},
major x tick style = transparent,
ybar=2*\pgflinewidth,
bar width=20pt,
ymajorgrids,
xmajorgrids,
xlabel style={font=\color{white!15!black}, font = \Large},
xlabel={Sparsity Levels},
ylabel style={font=\color{white!15!black}, font = \Large},
ylabel={Time (in ms)},
symbolic x coords={70\%,80\%,90\%, 95\%},
xtick = data,
scaled y ticks = false,
enlarge x limits=0.25,
ymin=0,
ymax=180,
legend cell align=left,
legend style={at={(0.51,0.7)}, nodes={scale=1.7}, anchor=south west, legend cell align=left, align=left, draw=white!15!black}
    ]
    \addplot[style={fill=mycolor1,mark=none}]
            coordinates {(70\%, 173) (80\%,111) (90\%,67) (95\%,38)};
\addlegendentry{RS Coding}
    \addplot[style={fill=mycolor3,mark=none}]
             coordinates {(70\%,101) (80\%,57) (90\%,37) (95\%,27)};
             \addlegendentry{Proposed Method}
    \end{axis}
\end{tikzpicture}
}
\caption{\small Comparison between our proposed method and RS coding based method in terms of computation time needed by a worker}
\label{fig3}
\end{figure}
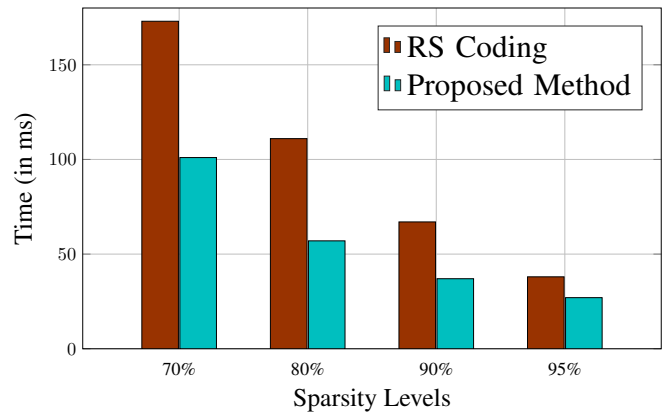

Fig. \ref{fig3} shows the maximum time needed by any worker for different approaches at different sparsity levels. It is evident that the workers take significantly less amount of time in the proposed method in comparison to the RS based approach. This is attributed to the fact that the sparsity level in the coded jobs can be preserved more in our proposed scheme than RS approach. For example, in our experiment, when matrix $\bfA$ has $90\%$ sparsity, the submatrices assigned to any worker in RS coding approach has around $51\%$ sparsity, whereas in our proposed scheme, even in the worst case, a parity worker can enjoy, on average $70\%$ sparsity. It should be noted that the message workers take further less time since they preserve the same level of sparsity as matrix $\bfA$.


\section{Conclusion}
In this paper, we present an approach for embedding distributed matrix-vector multiplication into the class of cross-parity check convolutional codes towards the goal of straggler mitigation. Our proposed scheme has significant advantages over RS based approaches. The recovery of the intended product is performed
using a low-complexity peeling decoder in our scheme as compared to polynomial interpolation in RS based approaches. Unlike RS based approaches which suffer from ill-conditioned recovery matrices, our recovery is numerically quite stable. Finally, for the case of sparse $\bfA$ matrices, our scheme requires much sparser combinations of the block-rows of $\bfA$, leading to faster computation at the worker nodes. Numerical examples have confirmed these results.
\bibliographystyle{IEEEtran}
\bibliography{citations}

\begin{appendices}
\section{}
\label{A}
Now we are going to discuss about the coefficients in the $CP(n,k)$ schemes for $s = 2$ or $3$.

\textbf{Case $s = 2$:} In this case, from \eqref{eqZ} we can write
\begin{align*}
Z_{ij} = \; - \; \prod_{\ell = 0, \ell \neq j}^{1} \; \frac{D^{2+i} - D^{\ell}}{D^j - D^{\ell}} \; .
\end{align*} If $j = 0$, then we get
\begin{align*}
Z_{i0} = \; - \; \frac{D^{2+i} - D}{1 - D} = \; D^{i+1} + D^i + \dots + D^2 + D \; .
\end{align*} And if $j = 1$, then we get
\begin{align*}
Z_{i1} = \; - \; \frac{D^{2+i} - 1}{D - 1} = \;- \left( D^{i+1} + D^i + \dots + D + 1 \right) \; .
\end{align*} So, we can say the coefficients are $0$ or $\pm 1$.

\textbf{Case $s = 3$:} In this case, from \eqref{eqZ} we can write
\begin{align*}
Z_{ij} = \; - \; \prod_{\ell = 0, \ell \neq j}^{2} \; \frac{D^{3+i} - D^{\ell}}{D^j - D^{\ell}} \; .
\end{align*} If $j = 0$, then we can write
\begin{align*}
Z_{i0} = \; - \; D^3 \;\frac{\left(D^{2+i} - 1\right) \left(D^{1+i} - 1\right)}{(1 - D)(1 - D^2)} \; .
\end{align*} Here, we can have two different cases. If $i$ is even, then
\begin{align*}
Z_{i0} = \; - \; D^3 \; \left(\sum\limits_{r=0}^{\frac{i}{2}} D^{2r} \right) \left(\sum\limits_{r=0}^{i} D^{r} \right) .
\end{align*}On the other hand, if $i$ is odd, then
\begin{align*}
Z_{i0} = - D^3 \left(\sum\limits_{r=0}^{i+1} D^{r} \right)\left(\sum\limits_{r=0}^{\frac{i-1}{2}} D^{2r} \right) .
\end{align*}Thus after multiplication, in terms of absolute value, the highest coefficient would be $\Bigl\lfloor\dfrac{i}{2}\Bigr\rfloor + 1$ in both cases. It will also be maximum when $i$ is maximum, $k-1$, and that coefficient would $\Bigl\lfloor\dfrac{k-1}{2}\Bigr\rfloor + 1$. Next, if $j = 1$, then we can write
\begin{align*}
Z_{i1} = \; - \; \frac{\left(D^{3+i} - 1\right) \left(D^{3+i} - D^2\right)}{(D - 1)(D - D^2)}
\end{align*} which leads to
\begin{align*}
Z_{i1} = \left( D^{i+2} + D^{i+1} + \dots + D + 1 \right) \left( D^{i+1} + D^i + \dots + D \right) \; .
\end{align*} Thus after multiplication, the maximum coefficient would be $i+1$, and so, the maximum coefficient is $k$. Finally, if $j = 2$, then we can write
\begin{align*}
Z_{i2} = \; - \; \frac{\left(D^{3+i} - 1\right) \left(D^{2+i} - 1\right)}{(D^2 - 1)(D - 1)} \; .
\end{align*} Here, we can have two different cases again. If $i$ is even, then
\begin{align*}
Z_{i2} = \; - \; \left( D^{i+2} + D^{i+1} + \dots + 1 \right) \left( D^i + D^{i-2} + \dots + 1\right) \; .
\end{align*}On the other hand, if $i$ is odd, then
\begin{align*}
Z_{i2} = \; - \; \left( D^{i+1} + D^{i-1} + \dots + 1\right) \left( D^{i+1} + D^{i} + \dots + 1 \right) \; .
\end{align*} Thus after multiplication, in terms of absolute value, the highest coefficient would be $\Bigl\lfloor\dfrac{i+1}{2}\Bigr\rfloor + 1 $ in both cases. It will also be maximum when $i$ is maximum, $k-1$, and that coefficient would $\Bigl\lfloor\dfrac{k}{2}\Bigr\rfloor + 1$.  Thus the result follows.

\section{}
\label{B}
\begin{figure}[t]
\centering
\resizebox{0.9\linewidth}{!}{
\begin{tikzpicture}[auto, thick, node distance=2cm, >=triangle 45]

\draw
	node [block6, minimum height = 14em,fill=green!30] at (0,0.34) (b0) {}
    node [block6, below = 0.0005cm of b0,minimum height = 2em, pattern=north west lines, pattern color=black] (b4) {}
    node [block6, minimum height = 6em,fill=green!30] at (1.5,1.75) (b1) {}
    node [block6, below = 0.0005cm of b1,minimum height = 2em, pattern=north west lines, pattern color=black] (b5) {}
    node [block6, below = 0.0005cm of b5,minimum height = 8em, fill=red!30] (b8) {}
    node [block6, minimum height = 2em,fill=green!30] at (3,2.45)(b2) {}
    node [block6, below = 0.0005cm of b2, minimum height = 2em, pattern=north west lines, pattern color=black] (b6) {}
    node [block6, below = 0.0005cm of b6,minimum height = 12em, fill=red!30] (b9) {}
	node [block6, minimum height = 2em, fill=green!30] at (4.5,2.45) (b3) {}
	node [block6, below = 0.0005cm of b3, minimum height = 14em, fill=red!30] (b10) {}
	node [block6, minimum height = 16em,fill=red!30] at (6,0.0) (b7) {}
	
	node [] at (0,-3.2) (t0) {$t_0$}
    node [] at (1.5,-3.2) (t1) {$t_1$}
    node [] at (3,-3.2) (t2) {$t_2$}
    node [] at (4.5,-3.2) (t3) {$t_3$}
    node [] at (6,-3.2) (t4) {$t_4$}
    node [block6, minimum height = 20em, minimum width = 24em] at (3,0) (b00) {}
    ;
\draw[color=blue,dotted,ultra thick] (-0.6,-3.3) -- (3.15,3.3);
\draw[color=red, dashed,ultra thick] (-0.6,-1.7) -- (4.6,3.3);
\draw[color=black,-,ultra thick] (-0.6,-0.2) -- (5.9,3.3);

\end{tikzpicture}
}
\caption{\small Recovering blocks from the stragglers, where the green and red parts indicate the decoded and undecoded blocks, respectively}
\label{lem}
\end{figure}
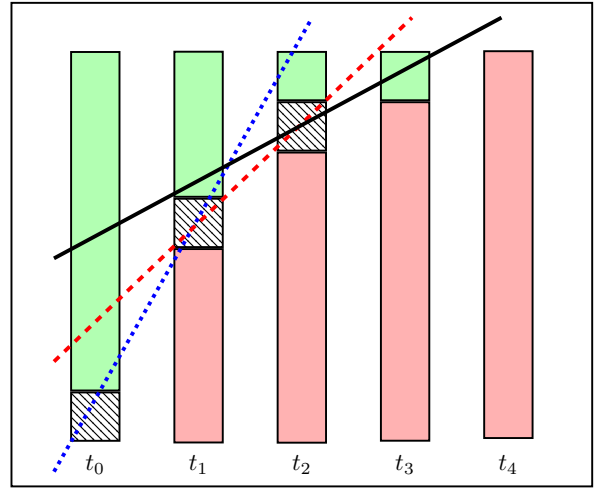
\begin{lemma}
We assume two stragglers having index $t_u$ and $t_v$, where $u < v$. We assume that two constraint lines with slopes $(s - 1 - u)$ and $(s - 1 - w)$ pass through the same point $c_{\alpha,t_u}$, where $w < u$. But, these two lines will pass through two different points, $c_{\beta,t_v}$ and $c_{\gamma,t_v}$ in worker $t_v$, since the slopes are not equal. Then it can be shown that
\begin{align*}
\gamma = \beta - (t_v - t_u)(u - w) .
\end{align*}
\end{lemma}
\begin{proof}
We can find the equation for the constraint line with slope $(s - 1 - u)$ from \eqref{eq1} as
\begin{align*}
\sum\limits_{j=0}^{n-1} c_{i - (s - 1 - u)j, j} = 0 ,
\end{align*} which passes through $c_{\alpha,t_u}$ and $c_{\beta,t_v}$, so we get
\begin{align*}
\alpha = \beta + (t_v - t_u)(s - 1 - u) .
\end{align*} And the constraint line with slope $(s - 1 - w)$ is given by
\begin{align*}
\sum\limits_{j=0}^{n-1} c_{i - (s - 1 - w)j, j} = 0 ,
\end{align*} which passes through $c_{\alpha,t_u}$ and $c_{\gamma,t_v}$, so we get
\begin{align*}
\alpha = \gamma + (t_v - t_u)(s - 1 - w) ,
\end{align*} which leads to
\begin{align*}
\gamma = \beta - (t_v - t_u)(u - w) .
\end{align*}
\end{proof}
From this lemma, we can say, if we have different constraint lines with different slopes passing through a particular point $\alpha$, in a straggler $t_u$, then the constraint line with a larger slope will pass through a prior point of a straggler $t_v$, if $u < v$, and vice versa.

We can see an example with five stragglers in Fig. \ref{lem}, where the green and red blocks indicate that they are decoded and undecoded, respectively. The slopes of the constraint lines are higher for decoding the straggler $t_0$ than for $t_1$. We assume that two different constraint lines from these two workers (blue for $t_0$ and red for $t_1$) intersect at a point in $t_1$. Then from the lemma we can say, the blue line will always intersect with the next stragglers prior to the red line does. And in a similar way, we can say that because of the larger slope, the red line will intersect with $t_3$ or $t_4$ before the black line; since these two lines have already intersected at a point in straggler  $t_2$.

\end{appendices}

\end{document}